\documentclass[12pt,twoside]{article}
\usepackage{fleqn,espcrc1}

\usepackage{graphicx}% Include figure files

\title{\Large Nuclear structure of $^{178}$Hf related to the spin-16, 31-year isomer}

\author{
Yang Sun,\address{Department of Physics, University of Notre Dame,
Notre Dame, Indiana 46556, U.S.A.}
\address{Department of Physics, Xuzhou Normal University, Xuzhou, 
Jiangsu 221009, P.R. China} 
Xian-Rong Zhou $^c$, Gui-Lu Long,\address{Department of Physics,
Tsinghua University, Beijing 100084, P.R. China}
En-Guang Zhao,\address{Institute of Theoretical Physics, Chinese
Academy of Sciences, PO Box 2735, Beijing 100080, P.R. China}
Philip M. Walker$^1$\address{Department of Physics, University of
Surrey, Guildford GU2 7XH, United Kingdom} }

\begin{document}
\maketitle

$^1$ Corersponding author: E-mail: p.walker@surrey.ac.uk;

phone: +44-1483-686807; fax: +44-1483-686781.

\bigskip
\begin{abstract}

ABSTRACT: The projected shell model is used to study the
multi-quasiparticle and collective excitations of $^{178}$Hf. With
an axially symmetric basis, the spin-16 isomer at 2.4 MeV appears
to be well separated in energy/spin space from other
configurations. However, projected energy surface calculations
suggest that $^{178}$Hf has significant softness to axially
asymmetric shapes, which can strongly modify the level
distribution. The implications for photodeexitation of the isomer
are discussed.

\end{abstract}

\bigskip



\bigskip

Long-lived isomers -- excited nuclear states with inhibited
electromagnetic decay -- may be considered to offer a form of
energy storage \cite{Ba97,Walker}. The possibility to trigger the
decay by the application of external electromagnetic radiation has
attracted much interest and potentially could lead to the
controlled release of nuclear energy. First reports of the
triggering of the spin-16, 2.4 MeV isomer in $^{178}$Hf with
low-energy ($<100$ keV) photons \cite{Col99,Col00,Col02} have,
however, been refuted \cite{Ahm01,Ahm03}. Meanwhile, the threshold
photon energy for triggering the isomer remains unknown.  This
situation contrasts with the recent determination of a threshold
triggering energy of 1.01 MeV for the spin-9, 75 keV isomer in
$^{180}$Ta \cite{Bel99}, and subsequently the discrete
intermediate states involved in the excitation/deexcitation
pathway have been identified \cite{Wal01}. The significantly
higher energy of the $^{178}$Hf isomer leads to the expectation of
a lower triggering threshold, on account of the higher level
density, even allowing for the pairing difference between the
even-even and odd-odd nucleon numbers in $^{178}$Hf and
$^{180}$Ta, respectively. The higher isomer energy in $^{178}$Hf
also leads to greater potential with regard to the utility of
triggered energy release.

\begin{figure}[htb]
\begin{minipage}[t]{150mm}
\includegraphics[scale=1.20]{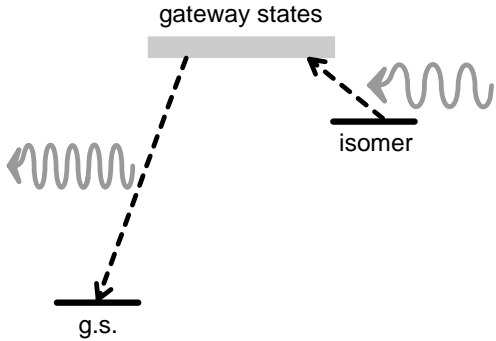}
\caption{Schematic diagram illustrating the process of triggered
gamma-ray emission from a nuclear isomer. }
\label{fig1}
\end{minipage}
\end{figure}

The two-step process of triggered gamma-ray emission is depicted
schematically in Fig. 1. A nucleus in an isomeric state is first
excited to an intermediate state by absorption of an incident
photon. An intermediate state may serve as a ``gateway'' that
connects the isomer to the ground state (g.s.). If the selection
rules for transitions between the gateway and the g.s. are
fulfilled, then enhanced gamma-decay, usually a multi-step
gamma-cascade, is expected to occur. In determining the favourable
conditions for triggering gamma-ray emission from the isomer, it
is necessary to have information on the structure of possible
gateway states, as well as possible paths of electromagnetic
transitions to and from these states. Experimentally, only a few
states close to the $^{178}$Hf isomer have been observed and
documented in the literature \cite{Mu97,Hay02}. The purpose of the
present letter is twofold: to demonstrate that the projected shell
model (PSM) \cite{review} is an appropriate theory for studying
high-spin isomers and associated excitations, including potential
gateway states; and to show that in the $^{178}$Hf case, in order
to trigger emission from the $16^+$ isomer with low-energy
photons, a series of external excitations via gateway states may
be necessary. In addition, the gateway states should be
sufficiently mixed with low-$K$ components.

The long half-life (31 years) of the 2.4 MeV isomer in $^{178}$Hf
is to a large extent dependent on the approximate conservation of
the $K$ quantum number, where $K$ is the projection of the angular
momentum on the body-fixed symmetry axis. The isomer has
$K^\pi=16^+$. Spontaneous electromagnetic transitions from a
high-$K$ state to a lower-energy, low-$K$ state are strongly
hindered by the $K$ selection rule \cite{BM75}, $|\Delta K| \leq
\lambda$, with $\lambda$ being the transition multipolarity. The
spontaneous decay of the $^{178}$Hf isomer involves inhibition
factors of close to 100 per degree of $K$ forbiddenness,
$\nu=\Delta K-\lambda$, with total inhibition $\approx$ $100^\nu$
\cite{Smi03}. This strong inhibition supports the view that
$^{178}$Hf has an axially symmetric intrinsic shape.

In a deformed, axially symmetric nucleus, a high-$K$ state is made
by summing the contributions from several unpaired quasiparticles.
An $n$-quasiparticle ($n$-qp) configuration gives rise to a
multiplet of $2^{n-1}$ states, with the total $K$ expressed by $K
= |K_1 \pm K_2 \pm \cdots \pm K_n|$, where $K_i$ is for an
individual neutron or proton. Among the $2^{n-1}$ states, the
state with the highest $K$ value, $K = \sum_{i}|K_i|$, is usually
energetically favoured and the most likely to form an $yrast$
state (lowest energy state for a given angular momentum). Deformed
nuclei with $A\approx 180$ have several high-$K$ single-particle
(both neutron and proton) orbitals close to the Fermi surface, and
several multi-qp high-$K$ states in this mass region are found to
be yrast \cite{Voi83,Wal01a}.

The physics of multi-qp states is well incorporated in the PSM
framework. The PSM follows closely the shell-model philosophy, and
in fact is a shell model constructed in a deformed multi-qp basis.
More precisely, the basis is first built in the qp basis with
respect to the deformed-BCS vacuum; then rotational symmetry,
violated in the deformed basis, is restored by angular-momentum
projection \cite{RS80} to form a rotational-invariant basis in the
laboratory frame; finally a two-body Hamiltonian is diagonalised
in the projected basis. In contrast to mean-field methods
\cite{Frau01} employed in the study of multi-qp states, the PSM
can produce fully correlated shell-model states, and can generate
well-defined wave functions, allowing us to compute, without
further approximations, the quantities such as transition
probabilities.

Starting from the deformed Nilsson scheme \cite{NKM} plus a
subsequent BCS calculation, one builds the shell model space (for
even-even systems) through multi-qp states:
\begin{equation}
\left|\phi_\kappa\right> =
\left\{\left|0 \right>,
\alpha^\dagger_{n_i} \alpha^\dagger_{n_j} \left|0 \right>,
\alpha^\dagger_{p_k} \alpha^\dagger_{p_l} \left|0 \right>,
\alpha^\dagger_{n_i} \alpha^\dagger_{n_j} \alpha^\dagger_{p_k}
\alpha^\dagger_{p_l} \left|0 \right>, \cdots \right\} ,
\label{baset}
\end{equation}
where $\alpha^\dagger$ is the creation operator for a qp and the
index $n$ ($p$) denotes neutron (proton) Nilsson quantum numbers
which run over the low-lying orbitals. The qp states are defined
in a space with three major shells ($N=4,5,$ 6 for neutrons and
$N=3,4,$ 5 for protons). The corresponding qp vacuum is $\left|0
\right>$. The indices $n$ and $p$ in (\ref{baset}) are general;
for example, a 2-qp state can be of positive (or negative) parity
if both quasiparticles $i$ and $j$ are from the same (or two
neighbouring) major shell(s). Positive and negative parity states
span the entire configuration space with the corresponding matrix
in a block-diagonal form classified by parity. Although in the
present work we truncate the model space beyond 4-qp's, the bases
(\ref{baset}) can be easily enlarged by including higher-order
multi-qp states. If the configurations denoted by ``$\cdots$" are
completely included, one recovers the full shell-model space
written in the representation of qp excitations.

In the case that the nuclear potential is axially symmetric, the
basis states in (\ref{baset}) are labelled by $K$. Thus, the {\it
projected} multi-qp $K$-states are the {\it building blocks} in
our shell-model wave function
\begin{equation}
\left|\Psi^{I\sigma}_M\right> = \sum_\kappa f^{I\sigma}_\kappa
\hat P^I_{MK_\kappa} \left|\phi_\kappa\right> .
\label{wavef}
\end{equation}
In Eq. (\ref{wavef}), $\kappa$ labels the basis states and
$\sigma$ the states with the same angular momentum. $\hat
P^I_{MK}$ is the angular momentum projector \cite{RS80}. The
coefficients $f_{\kappa}^{I\sigma}$ in Eq. (\ref{wavef}) are
determined by diagonalisation of the Hamiltonian. Diagonalisation
is the process of configuration mixing (here, $K$-mixing). Thus,
the required physics, {\it i.e.} constructing multi-qp states
(with good angular momentum and parity) and mixing these states
through residual interactions, is properly incorporated in the
model. Electromagnetic transition probabilities can be directly
computed by using the resulting wave functions.

The Hamiltonian employed in the PSM \cite{review} is
\begin{equation}
\hat H = \hat H_0 - {1 \over 2} \chi \sum_\mu \hat Q^\dagger_\mu
\hat Q^{}_\mu - G_M \hat P^\dagger \hat P - G_Q \sum_\mu \hat
P^\dagger_\mu\hat P^{}_\mu ,
\label{hamham}
\end{equation}
where $\hat H_0$ is the spherical single-particle Hamiltonian,
which contains a proper spin--orbit force \cite{NKM}. The other
terms in Eq. (\ref{hamham}) are quadrupole-quadrupole, and
monopole- and quadrupole-pairing interactions, respectively. The
strength of the quadrupole-quadrupole force $\chi$ is determined
in such a way that it has a self-consistent relation with the
quadrupole deformation $\varepsilon_2$. The monopole-pairing force
constants $G_M$ are
\begin{equation}
\begin{array}{c}
G_M = \left[ 20.12 \mp 13.13 \frac{N-Z}{A}\right] ~A^{-1} ,
\label{GMONO}
\end{array}
\end{equation}
with ``$-$" for neutrons and ``$+$" for protons, which reproduces
the observed odd--even mass differences in the mass region.
Finally, the strength parameter $G_Q$ for quadrupole pairing was
simply assumed to be proportional to $G_M$, with a proportionality
constant 0.16, as commonly used in PSM calculations \cite{review}.

\begin{figure}[htb]
\begin{minipage}[t]{150mm}
\includegraphics[scale=1.2]{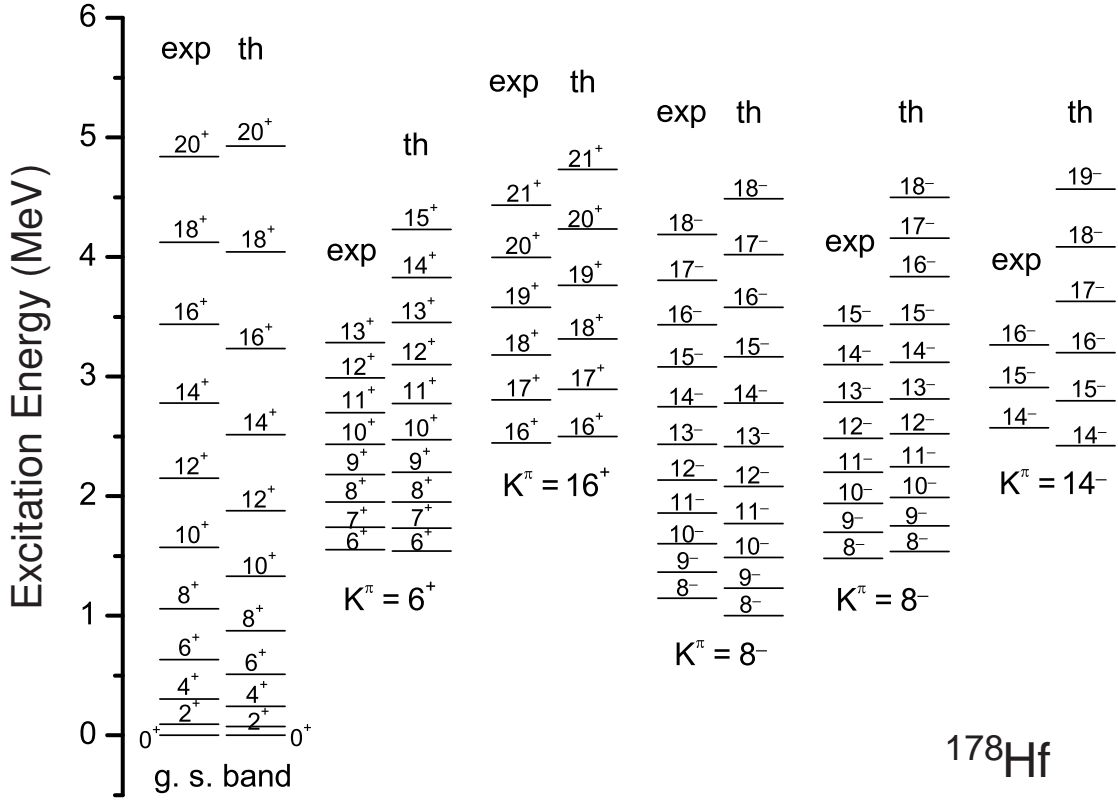}
\caption{Comparison of calculated energy levels in $^{178}$Hf with
known data \protect\cite{Mu97}.}
\label{fig2}
\end{minipage}
\end{figure}

For $^{178}$Hf, the model basis is built with the deformation
parameters $\varepsilon_2=0.251$ and $\varepsilon_4=0.056$. These
values are taken from the literature \cite{BFM86}. Fig. 2 shows
the calculated energy levels in $^{178}$Hf, compared with the
known data \cite{Mu97}. Satisfactory agreement is achieved for
most of the states, except that for the bandhead of the first
$8^-$ band and the $14^-$ band, the theoretical values are too
low. The dominant structure of each band can be read from the wave
functions. We found that the $6^+$ band has a 2-qp structure $\{
\nu [512]{5\over 2}^- \oplus \nu [514]{7\over 2}^- \}$, the $16^+$
band has a 4-qp structure $\{ \nu [514]{7\over 2}^- \oplus \nu
[624]{9\over 2}^+ \oplus \pi [404]{7\over 2}^+ \oplus \pi
[514]{9\over 2}^- \}$, the first (lower) $8^-$ band has a 2-qp
structure $\{ \nu [514]{7\over 2}^- \oplus \nu [624]{9\over 2}^+
\}$, the second (higher) $8^-$ band has a 2-qp structure $\{ \pi
[404]{7\over 2}^+ \oplus \pi [514]{9\over 2}^- \}$, and the $14^-$
band has a 4-qp structure $\{ \nu [512]{5\over 2}^- \oplus \nu
[514]{7\over 2}^- \oplus \pi [404]{7\over 2}^+ \oplus \pi
[514]{9\over 2}^- \}$. These states, together with many other
states (not shown in Fig. 2) obtained from the same
diagonalisation process, form a complete spectrum including the
high-$K$ isomeric states and candidate gateway states.

In order to study the possible gateway states through which
triggered isomeric emissions might occur, in Fig. 3 we plot the
high-$K$ bands lying close in energy to the $16^+$ isomer. We plot
also the calculated lower $8^-$ band via which the gamma-cascade
could reach the g.s. The $16^+$ isomeric band is found to be
yrast, above which there are many other 4-qp high-$K$ bands with
either positive or negative parity. It can be seen from Fig. 3
that the $16^+$ isomeric band is well separated from the other
bands, leaving an energy gap of nearly 500 keV. However, the
bandhead energies of the other high-$K$ bands are not much higher
than the $16^+$ isomer, and an external energy of $<$100 keV may
be able to excite the $16^+$ isomer.

A close look at Fig. 3 suggests, however, that in order to excite
the $16^+$ isomer to the states of the $8^-$ band, a series of
external excitations may be necessary, otherwise the deexcitation
would return to the original isomer. The excitation may proceed
stepwise among the gateway states, until it arrives in a low-$K$
state (a possible state is a member of the $8^-$ band) from which
spontaneous decay to the g.s. is possible. Of course, if such a
multi-step excitation is required, then the probability would be
very low. In addition, in order to link to the $8^-$ band, the
gateway states should contain sufficient $K\approx 8$ components.
The latter requirement is usually difficult to fulfill in an
axially symmetric nucleus; however, the discussion below may open
a possibility, through non-axial distortions.

\begin{figure}[htb]
\begin{minipage}[t]{150mm}
\includegraphics[scale=1.2]{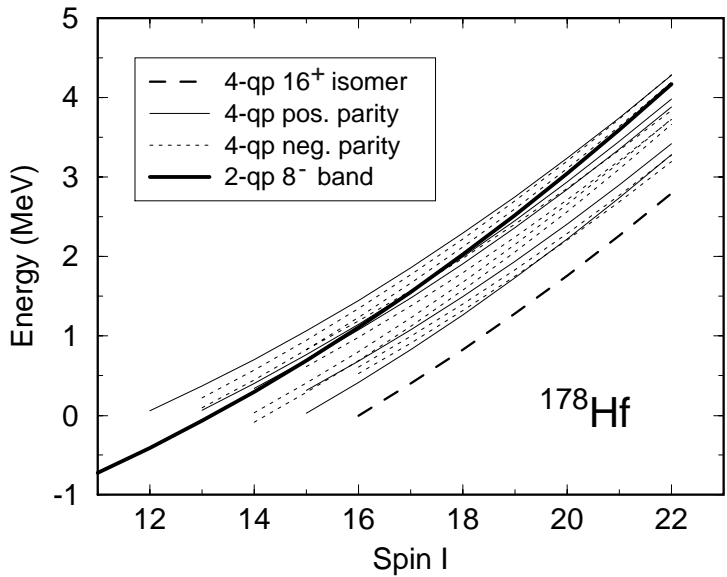}
\caption{Calculated high-$K$ bands in $^{178}$Hf. The bands are
plotted relative to the bandhead energy of the $16^+$ isomer.}
\label{fig3}
\end{minipage}
\end{figure}

The picture shown in Fig. 3 has been obtained by assuming an axial
basis, and it could be modified if the nucleus exhibits
significant deviation from the axial shape. The amount of triaxial
deformation is given by an additional deformation parameter,
$\gamma$. To see if the $\gamma$ degree of freedom plays a role in
$^{178}$Hf, we have newly extended the capability of the PSM to
study energy surfaces as a function of the two quadrupole
deformation parameters, $\varepsilon_2$ and $\gamma$, based on
exact projection calculations in three-dimensional Euler space
\cite{TPSM1,TPSM2}. Here, we calculate the
angular-momentum-projected energies having the form
\begin{equation}
E^I(\varepsilon_2,\gamma) = {{\left<
\phi(\varepsilon_2,\gamma)\right|\hat H \hat P^I \left|
\phi(\varepsilon_2,\gamma)\right>} \over {\left<
\phi(\varepsilon_2,\gamma)\right| \hat P^I \left|
\phi(\varepsilon_2,\gamma)\right>}}.
\label{energy}
\end{equation}
For comparison, the unprojected energy surfaces are also
calculated. The unprojected energy contains a mixture of states
with good angular momenta.

\begin{figure}[htb]
\begin{minipage}[t]{150mm}
\includegraphics[scale=1.2]{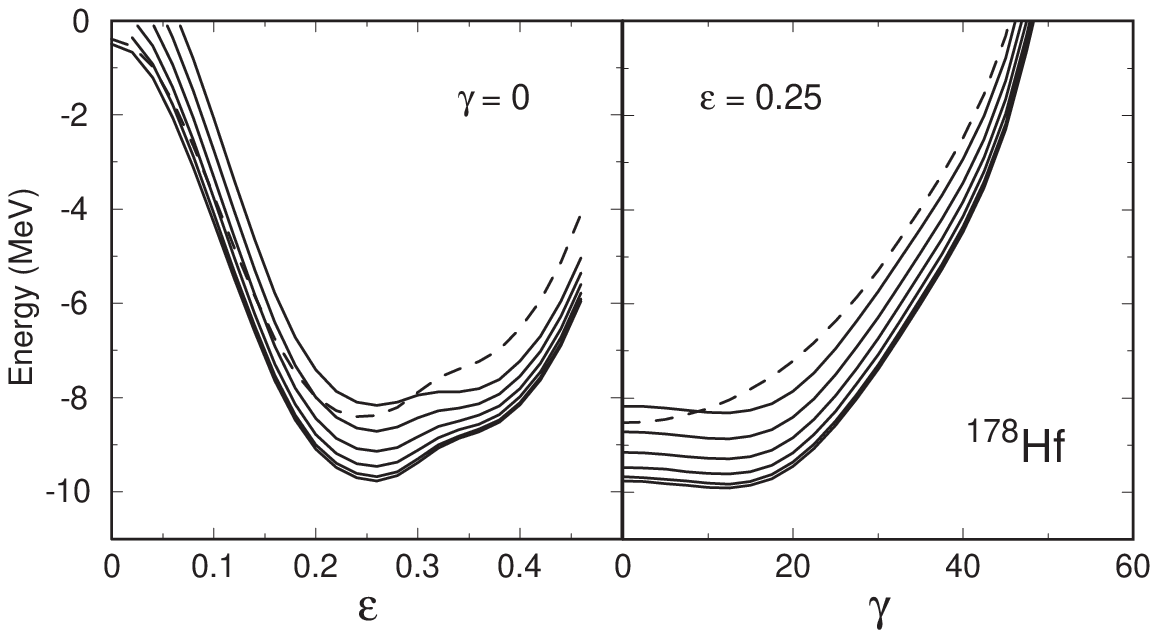}
\caption{Energy surfaces for states with angular momentum $I=0,
2,\cdots, 10$. The energies are calculated as function of (a)
quadrupole deformation $\varepsilon_2$ with $\gamma=0$, and (b)
$\gamma$ deformation with $\varepsilon_2=0.25$. Full (dashed)
curves correspond to projected (unprojected) calculations.}
\label{fig4}
\end{minipage}
\end{figure}

Energy surfaces for states with angular momentum $I=0, 2,\cdots,
10$ are plotted in Fig. 4. The left part of Fig. 4 shows the
energies as a function of $\varepsilon_2$ with $\gamma=0$. A
minimum at $\varepsilon_2=0.25$ is seen, which is consistent with
the use of $\varepsilon_2=0.251$ in the calculations of Figs. 2
and 3. The unprojected energy surface has a similar form, but the
minimum has a slightly smaller $\varepsilon_2$ value. In the right
part of Fig. 4, energies are calculated as a function of $\gamma$,
with the minimum value of $\varepsilon_2=0.25$. It is interesting
to observe that the unprojected surface has a minimum at
$\gamma=0$, suggesting that this nucleus is axially symmetric.
However, the projected surfaces exhibit a qualitatively different
pattern: For all the spin states with $I=0, 2,\cdots, 10$, the
energies with small $\gamma$ are almost constant, and they start
to rise only for $\gamma > 15^\circ$. This suggests a softness to
axially asymmetric shapes. Note that such qualitative differences
between projected and unprojected calculations have been discussed
before in another context \cite{HHR84}.

Nuclei having flat energy surfaces in the $\gamma$ direction have
recently been suggested \cite{Y5} to be critical point nuclei
sitting in the transitional region between the axially deformed
and triaxially deformed phases. The energy surfaces in Fig. 4
indicate that $^{178}$Hf has a character close to that of critical
point nuclei with considerable $\gamma$ softness. Therefore, the
$\gamma$ degree of freedom may play a significant role in this
nucleus. Xu {\it et al.} \cite{Xu98} studied the importance of
$\gamma$ deformation for high-$K$ bandheads in $^{178}$W and
$^{182}$Os, and for some hafnium isotopes \cite{Xu00}. While axial
symmetry was predicted for the hafnium cases, the calculations for
multi-qp states were restricted to bandhead shapes and energies,
and not the associated rotational excitations. Our calculations
now suggest that $^{178}$Hf exhibits a significant $\gamma$
softness already near its ground state. The $\gamma$ softness may
strongly enhance excitation of the gateway states. However, an
extension of the PSM for the description of multi-qp states, with
the $\gamma$ degree of freedom explicitly included, does not yet
exist.

To summarise, in the PSM the multi-qp states with good angular
momentum serve as the building blocks, and mixing of these states
is incorporated through two-body residual interactions. Our
calculations for $^{178}$Hf show that the $16^+$ isomer lies well
separated in energy from other states, suggesting that a series of
external excitations may be necessary to trigger isomeric
emission. However, axially asymmetric energy-surface calculations
show that in $^{178}$Hf there is considerable $\gamma$ softness.
This fact, in the context of recent discussions of critical point
nuclei \cite{Y5}, indicates that not only this particular example,
but also other nuclei in this mass region, may be significantly
influenced by the $\gamma$ degree of freedom. The
$\gamma$-softness may strongly enhance excitation of the gateway
states, in favour of triggering isomer decay. To describe multi-qp
states properly, and to further guide experiments in finding
candidate nuclei for triggered gamma-ray emission, an advanced
theory following the projected shell model concept, with the
$\gamma$ degree of freedom included, is very much desired. Work
along these lines is in progress.

Y.S. thanks Professor D.-H. Feng for communications. Research at
University of Notre Dame is supported by the NSF under contract
number PHY-0140324.

\baselineskip = 14pt
\bibliographystyle{unsrt}

\end{document}